# The Search for Living Worlds and the Connection to Our Cosmic Origins

*Contact person:* **Prof Martin Barstow**

*Address:* **Department of Physics & Astronomy, University of Leicester, UK**
*Email:* **mab@le.ac.uk**
*Telephone:* **+44 116 252 3492**

*Imaging Earth 2.0. Simulation of the inner solar system viewed at visible wavelengths from a distance of 13 parsec with a LUVOIR telescope. The enormous glare from the central star has been suppressed with a coronagraph so the faint planets can be seen. The atmosphere of each planet can be probed with spectra to reveal its composition. Credit: L. Pueyo, M M. N'Diaye (STScI)/A. Roberge (NASA GSFC).*

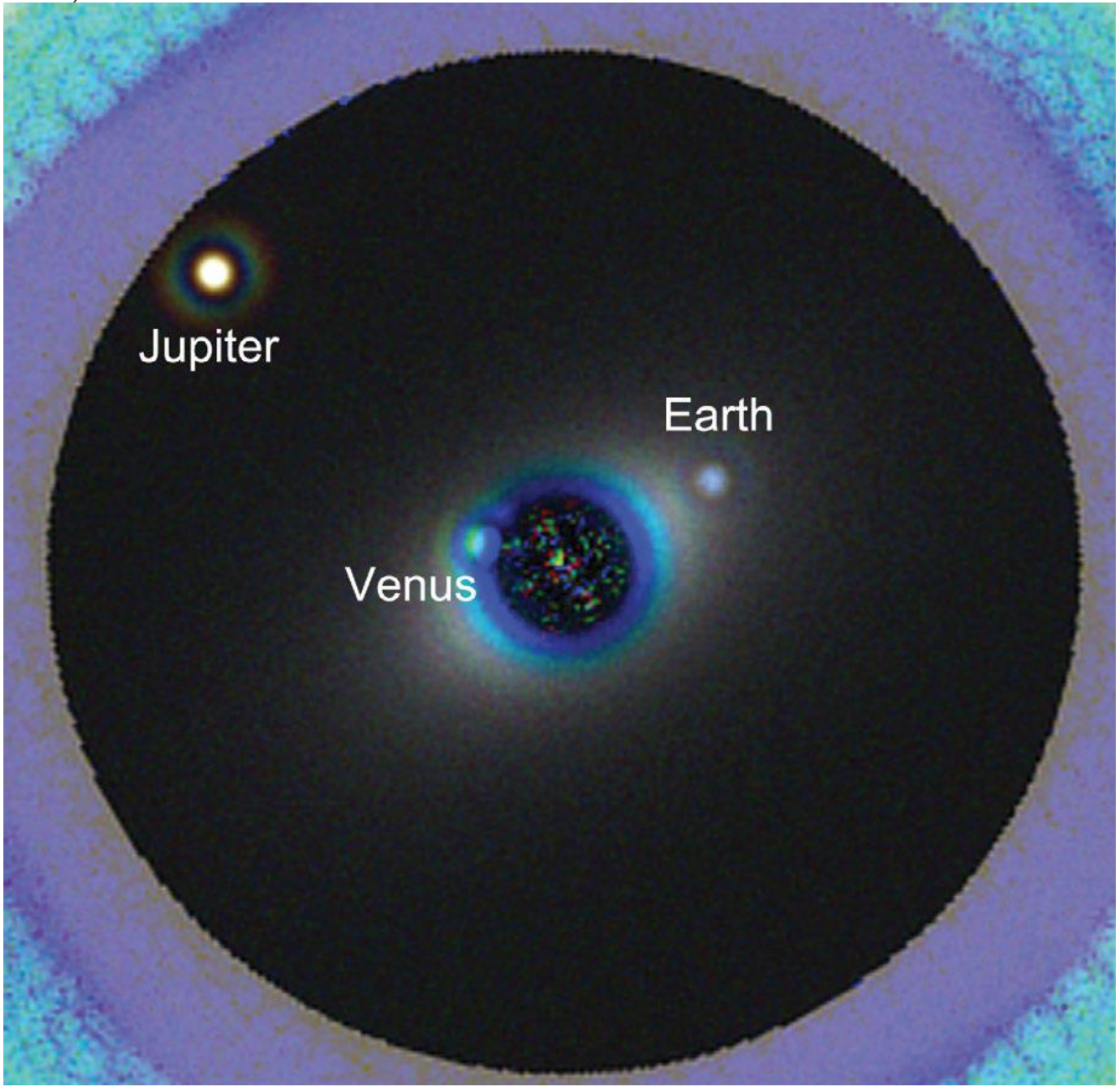

# Executive summary

One of the most exciting scientific challenges is to detect Earth-like planets in the habitable zones of other stars in the galaxy and search for evidence of life.

During the past 20 years the detection of exoplanets, orbiting stars beyond our own has moved from science fiction to science fact. From the first handful of gas giants, found through radial velocity studies, detection techniques have increased in sensitivity, finding smaller planets and diverse multi-planet systems. Through enhanced ground-based spectroscopic observations, transit detection techniques and the enormous productivity of the Kepler space mission, the number of confirmed planets has increased to more than 2000. There are several space missions, such as TESS (NASA), now operational, and PLATO (ESA), which will extend the parameter space for exoplanet discovery towards the regime of rocky earth-like planets and take the census of such bodies in the neighbourhood of the Solar System. The ARIEL mission (ESA) will also be able to characterize the atmospheres of gas- and ice-giants.

The ability to observe and characterise dozens of potentially Earth-like planets now lies within the realm of possibility due to rapid advances in key space and imaging technologies. The associated challenge of directly imaging very faint planets in orbit around nearby very bright stars is now well understood, with the key instrumentation also being perfected and developed. Such advances will allow us to develop large transformative telescopes, covering a broad UV-optical-IR spectral range, which can carry out the detailed research programmes designed to answer the questions we wish to answer:
- Carry out high contrast imaging surveys of nearby stars to search for planets within their habitable zones.
- Characterise the planets detected to determine masses and radii from photometric measurements.
- Through spectroscopic studies of their atmospheres and surfaces, search for habitability indicators and for signs of an environment that has been modified by the presence of life.

Active studies of potential missions have been underway for a number of years. The latest of these is the Large UV Optical IR space telescope (LUVOIR), one of four flagship mission studies commissioned by NASA in support of the 2020 US Decadal Survey. LUVOIR, if selected, will be of interest to a wide scientific community and will be the only telescope capable of searching for and characterizing a sufficient number of exoEarths to provide a meaningful answer to the question "Are we alone". Therefore, we propose that **ESA should become an international partner in LUVOIR, contributing at the level of an M-class mission or above.**

# 1. Introduction
One of the most exciting scientific discoveries of the past 20 years has been the detection of exoplanets, orbiting stars beyond our own. From the first handful of gas giants, found through radial velocity studies, detection techniques have increased in sensitivity, finding smaller planets and diverse multi-planet systems. Through enhanced ground-based spectroscopic observations, transit detection techniques and the enormous productivity of the Kepler space mission, the number of confirmed planets has increased to more than 2000. As techniques have improved, an amazing diversity of planet and system types has emerged. A tremendous, previously unanticipated, advance is the ability to characterise the atmospheres of giant exoplanets through spectroscopic observations made in and out of transit. HST has been at the forefront of this work and it is anticipated that JWST will make further breakthroughs. However, neither instrument is capable of direct studies of exoEarths. There are several space missions, such as TESS (NASA) and PLATO (ESA), which will extend the parameter space for exoplanet discovery towards the regime of rocky earth-like planets and take the census of such bodies in the neighbourhood of the Solar System. However, at the moment, one of the major scientific goals of exoplanet research lies tantalisingly beyond the capabilities of the current suite of space missions and ground-based telescopes: Are we alone in the Universe?  To answer this question requires the construction of a large UVOIR space-based observatory. With the long development timescales required to bring such a mission to fruition, there is an urgent need to plan for such a mission. An earliest launch date in the mid-2030s aligns with the ESA Vision 2050.

The ability to observe and characterise dozens of potentially Earth-like planets is within the realm of possibility due to rapid advances in key space and imaging technologies. The associated challenge of directly imaging very faint planets in orbit around nearby very bright stars is now well understood, with the key instrumentation also being perfected and developed. Such advances will allow us to develop large transformative telescopes, covering a broad UV-optical-IR spectral range.

For these exoplanet studies, the main scientific goals of a large UV-optical-IR telescope (LUVOIR) will be to:

- Carry out a high contrast imaging survey of nearby stars to search for planets within their habitable zones.
- Characterise the planets detected to determine masses and radii from photometric measurements.
- Through spectroscopic studies of their atmospheres and surfaces, search for habitability indicators and for signs of an environment that has been modified by the presence of life.

Achieving these goals requires access to a highly efficient optical system with a large collecting area, at a cost mirroring that of flagship missions such as HST and JWST, accounting for a large fraction of the science budgets of the lead agency and any major partners. Hence, a LUVOIR telescope is likely to be a multi-agency programme. The LUVOIR study exercise is already underway within NASA. While the key capabilities will be driven by exoplanet research LUVOIR will be transformational across many areas of astronomy and support the aspirations of the widest possible astronomical community, if operated as a general purpose observatory. For example, the facility will be capable of addressing ambitious goals from studying the earliest seeds of galaxies (>10 billion years ago), to the birth of galaxies like the Milky Way (6–10 billion years ago), to the birth of stars like the Sun in the Milky Way (5 billion years ago), culminating in the birth of planetary systems like our own (now). At each epoch, the radically sharp vision and sensitivity of a large space telescope will reveal things previously unseen, including how galaxies, stars, and planets play their part in establishing the conditions for life.

In the following sections we describe in detail the exoplanet science that could be undertaken with a facility like LUVOIR and summarise a selection of the wide range of other science goals than can be addressed along with the associated technical and instrument requirements.

## 2. Exoplanets

Planetary radiation consists mainly of thermally emitted radiation (at IR wavelengths) and of reflected light of the parent star (at UV – VIS – NIR wavelengths). Both types of radiation hold information on the composition and structure of the planet's atmosphere and/or surface. Traditionally, most instruments measure only the spectral dependence of the total flux (i.e. total energy) of planetary radiation. However, LUVOIR will measure both total and linearly polarized fluxes as functions of the wavelength.

The next decade will see a multiplication of space-based planet detection facilities (CHEOPS, TESS, and PLATO in particular [note that these missions have neither spectroscopic nor UV capabilities]) that will provide us with a large number of transiting planets covering the whole parameter space necessary to thoroughly study transiting exoplanet atmospheres and escape processes from an observational perspective. For example, the TESS mission is expected to find about 1700 planets orbiting nearby stars, some of which will lie in the habitable zone of M-dwarfs, while PLATO will greatly supersede these numbers, particularly by extending the search to longer-period planets (Sullivan et al. 2015; Rauer et al. 2014).

### 2.1 Direct imaging and spectroscopy of exoEarths

As of late 2016, ~20 exoplanets have been discovered with masses similar to or somewhat larger than the Earth in the "habitable zone" of their parent star (cf. Jenkins et al. 2015, i.e. at distances from their parent star that may allow liquid water on their surfaces), including a candidate planet in the habitable zone of our nearest stellar neighbor, Proxima Centauri b (Anglada-Escudé et al. 2016). However, whether or not these planets are truly habitable remains an open question – the habitable zone planets known to date have only been detected indirectly, and we cannot yet probe their atmospheric composition. Characterising the planetary context as well as the detection of biosignature gases in exoplanet atmospheres via spectroscopy and photometry is necessary to determine if so-called habitable zone planets are truly habitable or indeed, inhabited. There are only two methods – direct imaging and transit spectroscopy – that yield photons and hence spectra from planets, enabling direct measurement of atmospheric properties.

Such spectroscopic techniques are already in use today to characterise the atmospheres of extrasolar giant planets, mini-Neptunes, and super-Earths, all highly irradiated transiting planets (Sing et al. 2016, Kreidberg et al. 2014) as well as wide, young giant planets (Macintosh et al. 2015, Bonnefoy et al. 2016, Wagner et al. 2016). Exoplanet spectroscopy has revealed that clouds (formed of silicate condensates in the case of wide giant exoplanets) are likely common both for transiting and directly imaged planets (Kreidberg et al. 2014, Skemer et al. 2012, 2014.)

While transit spectroscopy is a highly valuable technique for characterizing exoplanets, due to viewing geometry, only ~10% of planets will have observable transits, and even less for habitable zone planets. Thus, direct imaging will provide a more complete census of the characteristics of planets, particularly those around higher mass stars where transits of habitable zone planets are too infrequent to characterise systems on reasonable timescales. Direct imaging spectroscopy of planets also literally provides a different angle compared to the currently dominant transmission spectrum method of studying transiting exoplanet atmospheres. For an Earth-like planet around a G2V star, refraction geometrical effects limit the depth of the atmosphere probed to 12 km and higher even in absence of clouds or hazes (Bétrémieux & Kaltenegger 2014). Many spectral features that are only marginally detected or undetected in transmission spectra would be easily detected in directly imaged spectra (Morley et al. 2015).

**A LUVOIR will be sensitive to almost all exoplanet atmospheres**, from hot to cold, close-in and far, gas giants to exo-Earths. Such sensitivity is necessary to properly place any exo-earths in context. It will also yield unprecedented new characterisation opportunities for higher mass planets. For instance, Crossfield et al. 2014 recently used the Doppler imaging technique to produce a surface map of the nearby brown dwarf Luhman 16B -- the first such map for any brown dwarf. With a high-resolution spectrograph and the high-contrast capabilities of LUVOIR, it will be possible to produce similar surface maps for directly imaged exo-Jupiters, allowing us to literally out map the cloud patterns on these planets (Crossfield 2014).

Habitable zone planets will likely be imaged in the next two decades -- Proxima Centauri b will be easily imaged by an extreme AO coronagraph at the E-ELT (Turbet et al. 2016) and the E-ELT instrument METIS may detect additional exo-Earths around somewhat higher mass nearby early M stars (Crossfield 2013). However, without spectroscopic confirmation of the presence of biosignature gases, such as $O_2/O_3$ in combination with $CH_4$, in context with other planetary indicators of habitability such as $H_2O$ and $CO_2$, it will be unclear if such planets are truly habitable or inhabited (Lovelock 1965).

Additionally, LUVOIR will enable UV observations (inaccessible from the ground), critical to understanding the photochemistry of these atmospheres. A low stellar UV environment allows certain biosignatures undetectable in an Earth-Sun analogue to potentially build up to detectable levels, such as $N_2O$ around quiescent M dwarfs (Rugheimer et al., 2015). As well the ratio of far-UV to near-UV is what determines the amount of $O_3$ in an exo-Earth atmosphere (Segura et al., 2005). Most of the known false positive mechanisms for $O_2$ and $O_3$ are informed in part by the UV environment of the host star (Domagal-Goldman et al., 2014; Luger and Barnes 2015; Harman et al., 2015). Thus it will be vital to have contemporaneous measurements of the UV stellar radiation field of habitable exoplanets in order to contextualize the atmospheric abundance of biosignatures in habitable exoplanets.

To build a picture of habitability as a function of host star, planet mass, etc., it is vital to characterise habitable zone planets orbiting a wide range of host stars. In Figure 1, contrasts achievable with current and future high-resolution imagers are plotted alongside current known directly imaged exoplanets and our own solar system planets (assuming a distance of 20 pc). There are 64 G0-G5 stars within 20 pc of the Earth. While earlier missions and telescopes such as JWST, WFIRST, and the ELTs will yield photometry and spectroscopy of a handful of habitable zone planets around low mass M and K dwarfs (e.g. Proxima Centauri b), only LUVOIR will yield the contrasts and resolutions necessary to detect and characterise true Earth twins (i.e. Earth-sized planets in the habitable zone of Sun-like stars). Through a LUVOIR-class mission we will be able to observe enough planets to reach statistically meaningful conclusions regarding the frequency and characteristics of exo-Earths and better understand our own planet in its astronomical context.

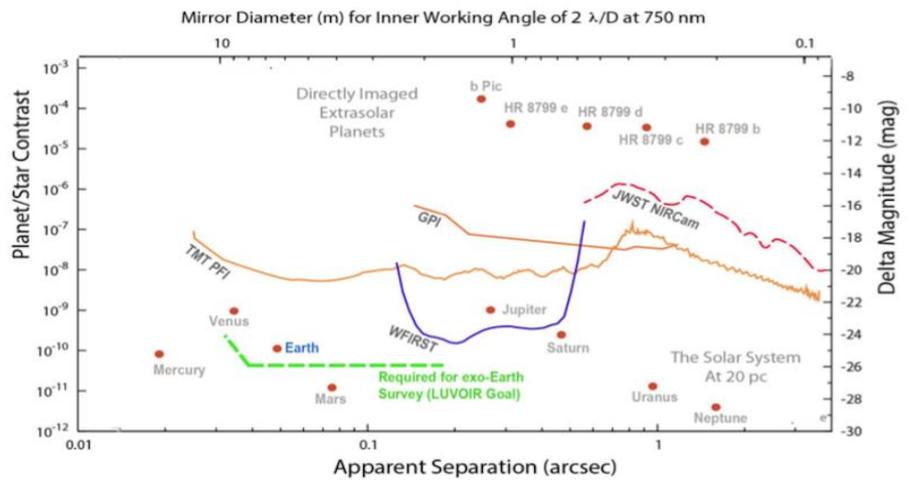

*Fig 1. Contrast limits achieved (S/N=5) in one hour's worth of observations (+post-processing) with various existing and planned coronagraphic instruments. Only LUVOIR will be sensitive to true Earth twins. From the NASA Goddard LUVOIR website, adapted from Lawson et al. (2012) and Mawet et al. 2012).*

Coronagraphic capabilities will also enable the study of Jupiter and Saturn analogues. Some observations of this class of planet (of order 10 targets) are likely to be obtained with the WFIRST spacecraft at wavelengths from 0.4 – 1 µm (Figure 2), but a larger platform would require shorter integration times and would therefore be able to cover a far greater number of targets. In addition, the increased wavelength range into the near-IR and UV would enable further constraints to be placed on atmospheric properties. Space-based observations of Solar System gas giants from 0.2 – 5 µm in both reflected and thermal (> 4 µm) radiation have uncovered details of cloud structure and ammonia/phosphine chemistry (e.g. Giles et al. 2015, Fletcher et al. 2011, Karkoschka & Tomasko 2005.) R~150 ultraviolet measurements of Jupiter using IUE in the 1980s (Gladstone & Yung 1983) provided constraints on abundances of trace hydrocarbon species, providing important information about carbon chemistry on the planet. A large space-based coronagraph with spectral coverage across a substantial part of this wavelength range would therefore be extremely valuable
.
## 2.2 Transiting exoplanet atmospheres

Since the first exoplanet transit was observed in 2002 our understanding of these exotic worlds has increased enormously, mostly thanks to the technique of transit spectroscopy. As an exoplanet passes in front of its star during transit, some of the starlight is filtered through the planet's atmosphere, and it reaches the observer bearing signatures of any absorbers and scattering particles that are present in the atmosphere (e.g. Sing et al. 2016). In addition, when the same planet is eclipsed by the star, the difference between the total fluxes outside of and during the eclipse enables detection of thermal radiation from the planet in the infrared, and reflected starlight in the visible and ultraviolet (Evans et al. 2013). For a few hot planets orbiting bright stars, phase-resolved spectroscopic observations over a whole orbit have provided longitudinal maps of atmospheric properties (e.g. Knutson et al., 2012; Stevenson et al. 2014). Taken together, these measurements can provide sufficient information to constrain atmospheric structure and composition (e.g. Line et al. 2014, Lee et al. 2014, Barstow et al. 2014), and thence to draw inferences about atmospheric chemistry and dynamics (e.g. Kataria et al. 2015).

The vast majority of high-quality transit, eclipse and phase curve observations have been obtained using the HST Space Telescope Imaging Spectrograph (STIS), in high resolution (R~50,000) modes, and Wide Field Camera 3 (WFC3) instruments, and the Spitzer InfraRed Array Camera (IRAC). Together, these instruments cover the wavelength region between 0.3 and 5 microns, with several gaps and variable spectral resolution (e.g. Sing et al. 2016). JWST will cover this spectral region from 0.6 microns onwards, but the eventual failure of HST will spell an end to space-based near-UV/visible light measurements for the foreseeable future.

The near-UV and visible spectral region is particularly critical for the study of cloud and haze properties of exoplanets in transmission spectra. The scattering efficiency for atmospheric aerosols, particularly for small particles, tends to increase at shorter wavelengths, making this spectral region ideal for detecting their presence and placing constraints on particle size and altitude. Condensation and photochemical haze formation are both critical atmospheric processes, and characterisation of aerosols is therefore an important aspect of atmospheric science, with links to atmospheric chemistry and structure. In addition, a poor understanding of aerosol properties can lead to misinterpretations of other spectral characteristics, as molecular absorption features in cloudy extended atmospheres can mimic absorption features in compact, high molecular weight atmospheres (e.g. Line & Parmentier 2016).

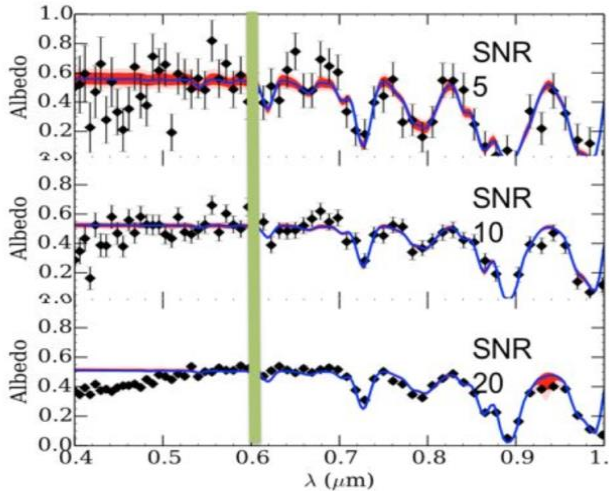

Figure 2: Simulations of Jupiter as seen in reflected light by the WFIRST coronagraph. Figure taken from Marley et al. (2014). Red and blue curves indicate model fits to the simulated data.

Only two examples exist so far of a reflection spectrum from secondary eclipse, but the geometric albedo measurement of HD 189733b in particular, showing substantial reflection at blue wavelengths, (Evans et al. 2013; Figure 3) is arguably one of the most important in the field. Determining planetary albedo, as well as measuring outgoing thermal flux, is necessary to understand the energy budget of a planet (Schwartz & Cowan 2015). The majority of known exoplanet host stars have significant output in the UV and visible. Hence, albedo measurements in these wavelength regions are necessary if the field is to progress. Such measurements are challenging as reflected light signatures are generally very small compared with the total system flux, and a space telescope with a large collecting area represents the only viable means of obtaining them. A UVOIR platform will allow reflected light and thermal signatures to be obtained with the same optics, minimising potential systematic problems with reconciling the data. Understanding energy deposition, recirculation and radiation in hot exoplanets requires an observatory of this kind.

As well as providing significant opportunities to advance the study of hot Jupiters, smaller temperate worlds will also be key targets for a large UVOIR observatory. The TRAPPIST-1 system (Gillon et al. 2016; 2017) hosts 7 transiting Earth-sized planets. TRAPPIST-1 planets are already key targets for JWST, but since the M8-type host star is red, small and faint, observations in the UV and visible with HST are more challenging. A larger platform would significantly enhance the likelihood of obtaining such measurements in the UV and visible, which would allow among other things a direct search for atmospheric oxygen as a biosignature through detection of the 0.75 μm $O_2$-A band. Characterising the atmospheres of nearby rocky, temperate worlds is a key goal for the immediate future of UVOIR astronomy, and a space-based spectroscopic platform with coronagraphic potential is a key part of this vision.

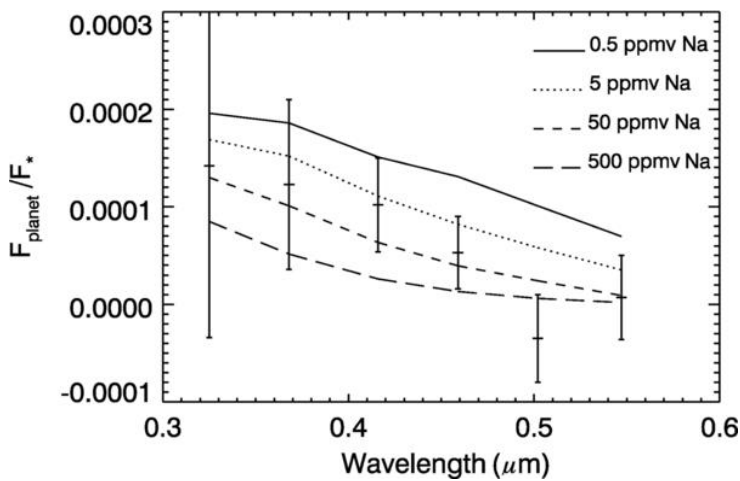

Figure 3: the reflection spectrum of hot Jupiter HD 189733b as observed by Evans et al. (2013) using HST/STIS. Figure and spectral model fits are taken from Barstow et al. (2014). We expect a 12 – 16 m class UVOIR platform to improve substantially on the quality of such spectra.

During transit, a planet may or may not occult star spots present on the star's surface. Transit spectra for planets orbiting stars with high spot coverage will contain signatures of occulted or unoccluded star spots, which can be misinterpreted as planetary atmospheric features. UV spectroscopic observations of the star prior to or after transit can help assessment of stellar activity, via measurement of CaII H and K lines. In addition, starspot contrast is strongest at shorter wavelengths, meaning that the presence of spots in lightcurves is more likely to be detected in the UV and enabling an estimate of the spot coverage. Stellar activity is a critical factor for planetary environment, especially for worlds in close orbits, and ultraviolet capability can provide key constraints in this regard.

Vidal-Madjar et al. (2003, 2004) obtained STIS far-ultraviolet transmission spectra of the close-in giant planet HD209458b revealing that the planet possesses an extended hydrogen atmosphere. How much of the extended atmosphere is hydrodynamically escaping (Koskinen et al. 2010) and how much is due to stellar wind interaction (Kislyakova et al. 2014) is still an open question (Ehrenreich et al. 2015). It is clear that atmospheric escape is a key factor shaping the evolution of close-in planets (e.g., Owen & Jackson 2012; Lopez & Fortney 2013) and their habitability (Lammer et al. 2009; Cockell et al. 2016), but the dependence of escape on the irradiating stellar UV flux and wind is far from being understood.

This atmospheric escape is driven mostly by the stellar X-ray and EUV photons that deposit energy (i.e., heat) in the planet upper atmosphere, but the majority of the stellar spectrum causing escape cannot be observed (the stellar EUV emission is absorbed by the interstellar medium for any star beyond a few pc). UV observations, particularly of Lyman alpha and other high formation temperature lines, allow one to infer the EUV stellar flux and hence correctly estimate the energy budget of an exoplanet atmosphere (Linsky et al. 2014; Chadney et al. 2015; Youngblood et al. 2016; France et al., 2018). By observing planetary transits at different wavelengths, from the FUV to the NUV, one also probes the intermediate region between the upper and lower atmosphere (Bourrier et al. 2015).

Atmospheric escape also has a major impact on our understanding of planet formation and consequently population (e.g., Kubyshkina et al., 2019). Because of high mass-loss rates, some low-mass close-in planets (e.g., CoRoT-7b, Kepler-10b) might be the remnant cores of evaporated Neptune-mass planets (Lecavelier des Etangs et al. 2004; Lecavelier des Etangs 2007; Leitzinger et al. 2011; Kurokawa & Kaltenegger 2013; Luger & Barnes, 2015; Kubyshkina et al., 2018). HST observations of the Neptune-size planet GJ436b showed that even warm giant planets can undergo significant escape and form very large tails of escaped planetary material (Fig. 4; Kulow et al. 2014; Ehrenreich et al. 2015; Bourrier et al. 2016). Escape processes, the upper atmosphere of planets, and its interactions with the host star can be thoroughly studied only at UV wavelengths, because at longer wavelengths the optical depth of the escaping planetary material is too low.

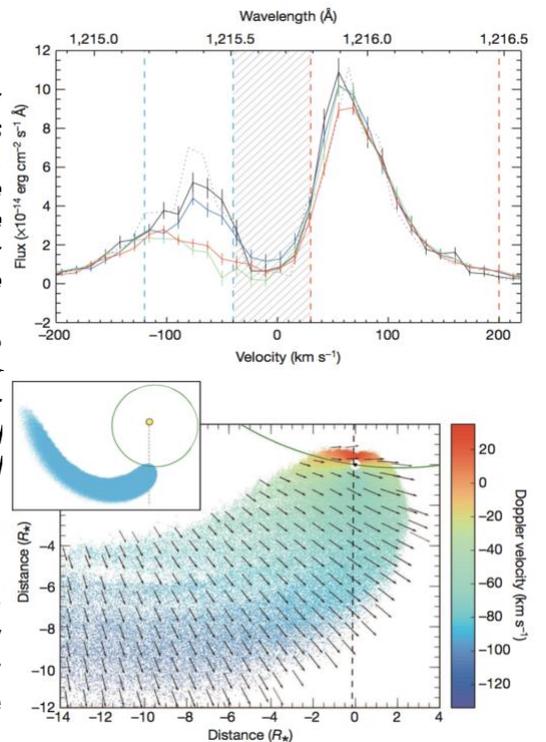

*Figure 4 – <u>Top</u>: average Lyman alpha line profiles of the M-type star GJ436 that hosts a warm Neptune-mass planet. The solid lines correspond to out-of-transit (black), pre-transit (blue), in-transit (green), and post-transit (red) observations from individual spectra. The line core (hatched region) cannot be observed from Earth because of the interstellar medium absorption along the line of sight. <u>Bottom</u>: polar view of a three-dimensional simulation representing a slice of the comet-like cloud coplanar with the line of sight of the GJ436 system. Hydrogen atom velocity and direction in the rest frame of the star are represented by arrows. Particles are color-coded as a function of their projected velocities on the line of sight (the dashed vertical line). Inset: zoom out of this image to the full spatial extent of the exospheric cloud (in blue). The planet orbit is shown to scale with the green ellipse and the star is represented with the yellow circle. From Ehrenreich et al. (2015).*

To fully explain the processes behind atmospheric escape requires the acquisition of UV transmission spectroscopy observations for about 150 transiting planets orbiting stars (later than spectral type A5) down to a V-band magnitude of 14. The brightest solar-like stars that are member of nearby, young (between 20 and 100 Myr), clusters and associations have V-band magnitudes between 12 and 14, thus the faint magnitude limit. The observations should ideally be carried out at both FUV and NUV wavelengths for the brightest/nearest systems (< 60 pc and/or brighter than V~9 mag, depending on the stellar characteristics) and only at NUV wavelengths for the remaining objects. A complete sample of planets is necessary to look for and study on a statistically significant basis, correlations of atmospheric escape with system parameters.

## 2.3 Spectropolarimetry of exoplanets

Polarimetry increases the planet/star contrast (Keller et al., 2010) and thus helps distinguishing planetary from stellar photons. Indeed, while direct starlight is virtually unpolarised (Kemp et al., 1987), starlight that has been reflected by a planet will generally be polarised, even when integrated across the planetary disk, and even for a symmetric planet (e.g. Seager et al., 2000; Stam et al., 2004). Measuring a polarised signal would directly confirm the planetary nature of an object. Indeed, the polarisation signal will be largest around planetary phase angles of 90° (that occur at least twice every planetary orbit), where the planet-star angular separation is largest. Possibly as a footnote: Note that a planet's thermal radiation will also be polarised when it is scattered by aerosol and clouds within a planet's atmosphere (de Kok et al., 2011). However, for a symmetric planet, the disk integrated thermal polarisation signal will be negligible, which is why we prefer polarimetry across the UVOIR.

The state of polarisation of the reflected starlight is more sensitive to the planet's physical properties, such as the cloud particles composition (see Fig. 5 & 6 from Karalidi et al., 2012; Rossi & Stam, 2017), than the total flux (see Hansen & Hovenier, 1974), especially when measured as functions of the phase angle and wavelength. Because the state of polarisation is a relative measure (polarised flux versus total flux), it is insensitive to e.g. distances and/or the planet's radius. Spectropolarimetry can thus play an essential role in characterising properties of planetary atmospheres and surfaces. For instance, the polarisation across the $O_2$ A band is strongly sensitive to cloud coverage and altitude (Fauchez & Stam, 2017). Polarised signals of exo-zodiacal dust disks and e.g. planetary rings will differ from those of exoplanets and allow the characterisation of planetary environments.

## 2.4 Detection and measurement of planetary magnetic fields

The presence of a planetary magnetic field appears to be an important ingredient for the advent and further evolution of complex life on the surface of a planet. A large scale magnetic field would screen the planetary surface and atmosphere from high-energy stellar wind particles that would lead to a quick escape of the atmosphere (Khodachenko et al. 2015), erode the atmosphere through ion pick-up from the stellar wind (hence the need to study also stellar winds; Kislyakova et al. 2014), and greatly damage the formation and stability of complex molecules, such as DNA. For a distant observer, Earth's magnetic field could be detected by measuring the bow-shock heading towards the Sun and lying at several Earth radii from the surface, well above Earth's atmosphere. This bow-shock is generated by the interaction of Earth's magnetic field with the supersonic solar wind.

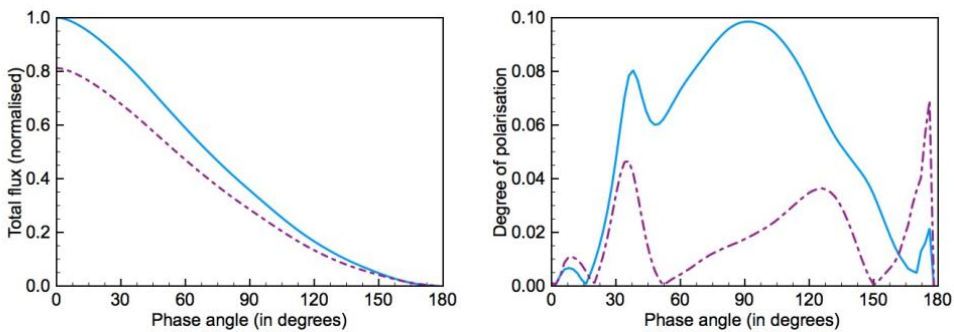

*Figure 5: Computed total fluxes (left) and degree of polarisation (right) of sunlight reflected by the Earth at 550 nm (solid, blue) and 865 nm (dashed, purple) as functions of the phase angle. The model Earth was covered with 63% liquid water clouds and 36% high-altitude ice clouds according to NASA/MODIS data taken on April 25$^{th}$, 2011. The polarisation maximum near 30 degrees phase angle is characteristic for liquid water cloud particles. The maximum around 90 degrees at 550 nm, is due to scattering by N2 and O2. At 865 nm, the polarisation is less sensitive to gaseous molecules. From Karalidi et al. (2012).*

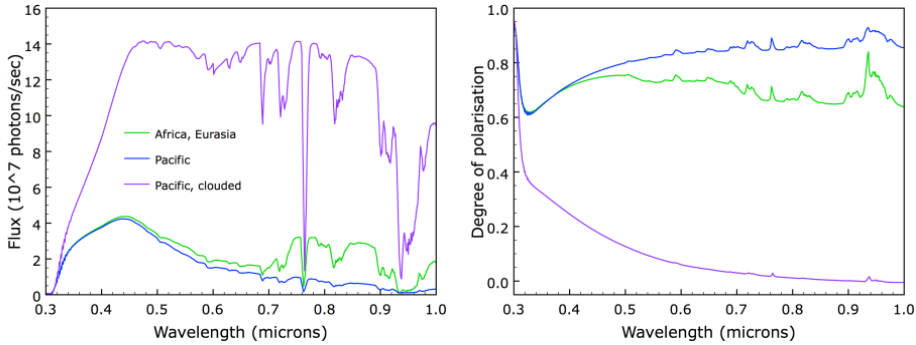

*Figure 6: Computed total fluxes (left) and degree of polarisation (right) of sunlight reflected by the Earth at a phase angle of 90 degrees, with different regions of the Earth turned towards the observer (from Karalidi et al., 2012).*

Transit asymmetries (i.e., early ingress; Fig. 7), some of which are believed to be caused by bow-shocks, have been detected at UV and optical wavelengths (e.g., Fossati et al. 2010a; Vidotto et al. 2010, 2011; Ben-Jaffel & Ballester 2013; Cauley et al. 2016). Bow-shocks could be of magnetic or non-magnetic origin and in some cases, from repeated observations and detailed modeling, it is possible to distinguish between them (Bisikalo et al. 2013; Turner et al. 2016). This can be done for example by measuring the distance between the planet and the head of the bow-shock from the timing of the early-ingress, where a large distance would hint at a magnetic support of the bow-shock, that could then be confirmed by follow-up modeling. Transit observations with LUVOIR will allow us to detect and measure transit asymmetries. Follow-up modeling will allow us then to infer their origin and in case planetary magnetic field strengths (Vidotto et al. 2010).

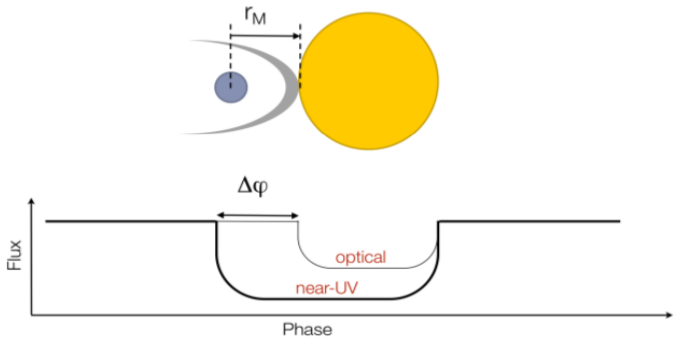

*Figure 7. Sketch of the light curves obtained through observations in the optical and in certain NUV lines, where the bow shock surrounding the planetary magnetosphere is also able to absorb stellar radiation. The stand-off distance from the shock to the center of the planet is assumed to trace the extent of the planetary magnetosphere $r_M$, which can be estimated by measuring the difference between the phases ($\Delta\varphi$) at which the NUV and the optical transits begin. From Vidotto et al. (2013).*

## 2.5 The bulk composition of rocky planetesimals and planets

Knowledge of the bulk composition of exoplanets is of key importance for their detailed characterisation, and a complete understanding of planet formation. However, traditional exoplanet observations yield only bulk densities (from mass and radius measurements obtained for transiting planets), and interior structures and composition based on these measurements are model-dependent and subject to degeneracies (Rogers & Seager 2010; Dorn et al. 2015. For the foreseeable future, FUV spectroscopy of white dwarfs accreting planetary debris remains the only way to directly and accurately measure the bulk abundances of rocky exoplanetary bodies.

Practically all known planet host stars, including the Sun, will evolve into white dwarfs, and many of their planets will survive (Veras et al. 2013). Planetesimals and planets are perturbed onto star-crossing orbits by dynamical interactions with more massive planets (Debes & Sigurdsson 2002; Veras & Gänsicke 2015), where they are eventually tidally disrupted and accreted by the white dwarf (Jura 2003, Vanderburg et al. 2015). In a pioneering paper, Zuckerman et al. (2007) showed that measuring the photospheric abundances of debris-polluted white dwarfs provides an unrivalled window into the bulk composition of exoplanets, in full analogy to meteorites informing us about the composition of the solar system. High-resolution (R>20,000) ultraviolet spectroscopy is fundamental for this work, as the wavelength range contains strong transitions of the rock-forming elements (Si, Fe, Mg, O), refractory lithophiles (Ca, Al, Ti), and in particular of volatile elements (C, N, P, S) that must be resolved to trace the formation region of the planetary material relative to the snow line. HST/COS spectroscopy corroborates the rocky, volatile-depleted nature of the disrupted planetesimals and planets (Jura et al. 2012), with a variety of bulk compositions similar to, if not exceeding, that seen among solar-system bodies (Gänsicke et al. 2012, see

The Search for Living Worlds – ESA Voyage 2050 White Paper

Fig. 8). These results have already informed recent models of planet formation (Carter-Bond et al. 2012). Of particular importance for the properties of planetary systems are the C/O and Mg/Si ratios. C/O ratios > 0.8 would result in a radically different setup from the solar system, with O-chemistry replaced by C-chemistry, which is discussed abundantly in the literature (e.g. Moriarty et al. 2014). The Mg/Si ratio determines the exact composition of silicates, which in turn has implications for planetary processes such as plate tectonics.

Global insight into the chemistry of planetary systems as a function of the mass and age of their host stars requires far-UV abundance studies of a substantial number of white dwarfs. Both the number of available targets, and the sensitivity of HST/COS currently limit progress. Over the next decade, Gaia will astrometrically identify ~$10^5$ white dwarfs, which will be followed up with ground-based multi-object spectrographs (WEAVE, 4MOST, DESI). Detection of Ca K 393nm will select 1000s of systems accreting planetary debris spanning the full parameter range in host/progenitor properties, providing a rich target sample for detailed LUVOIR far-ultraviolet bulk composition measurements.

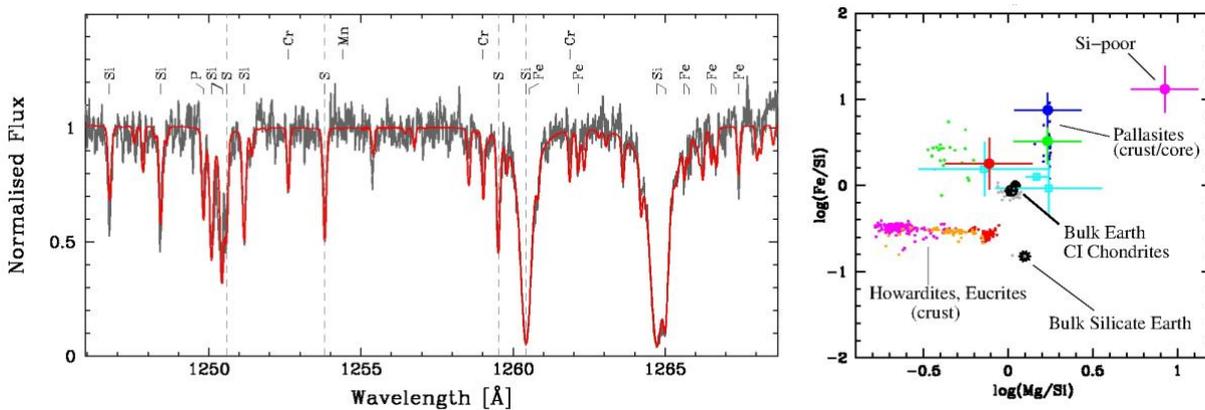

*Figure 8. Far-UV spectroscopy of white dwarfs accreting planetary debris (left, HST/COS), provides accurate bulk-abundances for rocky exo-planetary bodies with masses of ~$10^{20}$-$10^{25}$ g, analogous to meteorite studies tracing the chemistry of the solar system (right: large dots with error bars are extra-solar planetesimals, small dots are solar system meteorites).*

It is also possible to determine the abundances present by using transit spectroscopy of rocky surfaces of very close-in planets, such as Mercury. Such planets hold usually a mineral atmosphere that is composed of the minerals evaporated from the rocky solid or liquid (i.e., lava) surface (e.g., Rappaport et al. 2012; Demory et al. 2016). Minerals evaporate from the planetary surface under the action of the intense stellar wind (sputtering) and then escape from the planet (Fig. 9; e.g., Mura et al. 2011; Miguel et al. 2011; Ito et al. 2015; Kislyakova et al. 2016). UV and optical high spectral resolution transit observations would allow one to characterise the physical properties of the mineral atmosphere, including the identification of the major chemical species forming the atmosphere (e.g., Ca, Fe, Na, Si) and of their relative abundance.

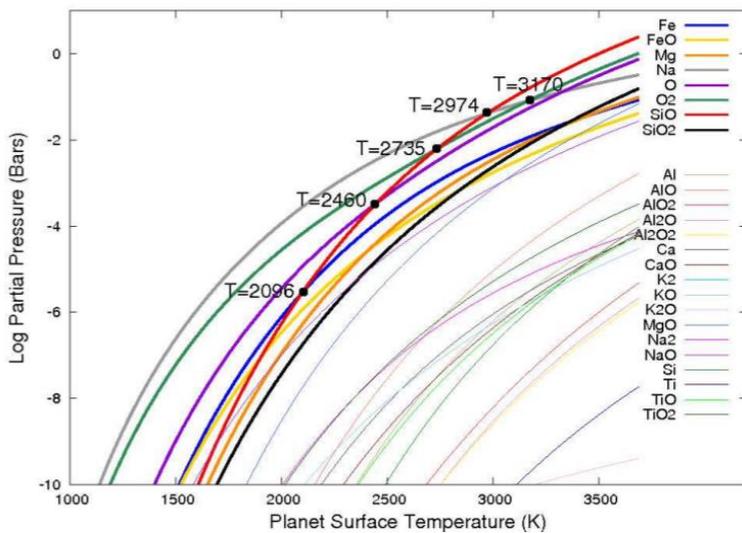

Figure 9. Planet surface temperatures vs. partial pressures of the gases vaporized from a komatiite magma. Temperatures at which the dominant gases change are indicated. From Miguel et al. (2011)

It is currently believed that several of these close-in rocky planets would be the remnant core of Neptune-mass planets that have lost their atmosphere through escape (e.g., Jin et al. 2014). If this is the case, the study of mineral atmospheres would allow us to infer the chemical composition of the core of gaseous giant planets, something which is not possible even for the solar system planets. The results obtained from such observations would be invaluable to inform planetary formation and structure models. The CHEOPS, TESS, and PLATO space missions will discover several of such close-in rocky planets, which recent observations (by CoRoT, Kepler, and K2) have shown constitute a non-negligible population of all planets in the galaxy.

## 3. The Solar System

The exploration of the solar system is fundamental to answering key questions in modern physical sciences. The ESA Cosmic Vision Themes 1 and 2 (What are the conditions for planet formation and the emergence of life?" and How does the Solar System work?") are directly and specifically related to the solar system, the understanding of which also interfaces directly with circumstellar disks and exoplanetary systems which represent the planet formation process at different stages. Similar themes are expected to be important for Voyage 2050. The various bodies in the solar system tell different portions of the solar system's formation story, the planets having formed by coalescence of planetesimals, of which evidence remains in the form of asteroids, comets, and Trans-Neptunian Objects (TNOs). Some of the key goals for solar system science follow:

### 3.1 Atmospheres and magnetospheres

A LUVOIR, particularly the UV capability, will revolutionise our understanding of the dynamics and composition of planetary and satellite atmospheres, particularly at the poorly-understood planets of Uranus and Neptune. Observations of the abundance and distribution of species such as H, $H_2$, $CO_2$, CO, and $H_2O$, along with many organics and aerosols will lead to an understanding of important source and loss processes, volcanism, aeronomy, atmospheric circulation and long term evolution of planetary atmospheres, all phenomena which relate to wider issues such as historical and present habitability, terrestrial anthropogenic climate change, and the nature of the pre-solar nebula. Cyrovolcanic and dust plumes, such as those seen at Enceladus, are potentially paradigm-shifting discoveries for small bodies such as satellites and TNOs. Through sensitive, high resolution imaging of planetary and satellite auroral emissions, this observatory will also impart a detailed understanding of all of the planets' magnetospheres, revealing the internal magnetic fields and thus internal structure and formation, along with information as to how energy and matter flow through the solar system. Short of sending dedicated spacecraft, this UV telescope is the only way to investigate the magnetospheres of the ice giants Uranus and Neptune. Jupiter's magnetosphere in particular acts as a readily observable analogue for more distant astrophysical bodies such as exoplanets, brown dwarfs and pulsars. Importantly, this observatory will not duplicate but instead perfectly complement ESA's L1 mission JUICE (Jupiter arrival 2030).

### 3.2 Surfaces and rings

Surface spectroscopy and imaging will provide information on the ice and non-ice condensible (e.g. organic) components of surface layers, revealing interior processes and surface-atmosphere interactions, with important implications for habitability at e.g. Europa. Long term observations of albedo maps will allow seasonal effects to be studied. Observations of H, ice, organics and other minor species in planetary ring systems will reveal their composition, formation- and life-times.

## 3.3 Small Bodies

Minor bodies (comets and asteroids) are of particular interest as they are left over building blocks from our own proto-planetary disc, and their composition can be used to place strong constraints on planet formation models. In addition to its uses for extrasolar planetary systems, LUVOIR will help us understand planet formation through the study of bodies in our own solar system. For example, the main water products in cometary comas, H, O, and OH, along with the CO Cameron bands, can be uniquely observed at high spatial resolution in the UV. These observations, along with high sensitivity detections of C, S, N, D/H, and rare gases beyond the snow line, will unveil the temperature and density of the pre-solar nebula in which the comets formed.

The ESA Rosetta mission has recently revealed that comet 67P contains abundant oxygen and nitrogen (Bieler et al 2015, Rubin et al 2015), meaning it must have formed slowly in a cold environment (Davidsson et al 2016). Comet spectra have many features in the UV, and a large UV/optical space telescope would advance the field primarily by being sensitive to much fainter comets. This would mean that the sort of study currently only possible for the rare 'great comets' like Hale-Bopp, for example measurement of isotopic ratios in H, C, O, N via very high-resolution spectroscopy, could be extended to more typical short period comets. This is important because isotopic ratios are very sensitive to the conditions (especially temperature) at the location in the disc where the comet formed; being able to make this measurement for many comets would test disc models (e.g. Bockelée-Morvan et al 2015). The increase in sensitivity relative to current instrumentation would also allow measurements of very low activity comets to be performed. One of the surprising early results from Rosetta was the revelation that electron impact chemistry (in addition to the previously studied photochemistry) is an important process in the coma of low activity comets (Feldman et al 2015); the physics of the interaction between weakly active comets and the solar wind is an area that needs to be investigated further, and requires sensitive observations of relatively faint bodies. Observations of $O^+$ and $CO^+$ will reveal the nature of the interaction of comets with the solar wind. This is of interest not just to better understand this physics, but because there is increasing evidence of buried water ice in unexpected places in the solar system (especially the so-called 'Main Belt comets' in the asteroid belt), and the line between weakly active comets and inactive asteroids is becoming increasingly blurred (e.g. Hsieh & Jewitt 2006, Jewitt et al 2015). The ability to detect very low levels of outgassing from such bodies is the most effective way to map where water (ice) can be found in our solar system, which is again an incredibly effective diagnostic of the original formation location of small bodies and their subsequent evolution, and therefore a good test of planet formation models.

A major objective will be to detect comet-like activity in TNOs, Main Belt asteroids, Trojans and Centaurs, testing models of thermal evolution at large heliocentric distances. High angular resolution albedo maps and observations of binarity of thousands of Main Belt asteroids will provide information on their composition, and thus source material, while many tens of bright TNOs will be fully characterised by these UV observations.

## 4 Stars and stellar populations

### 4.1 The evolution of protoplanetary disks (PPDs)

Accretion disks are very common structures in the Universe- they are the means for nature to channel mass accretion when the intrinsic angular momentum of the infalling gas is high. They work on a poorly known mechanism that transports angular momentum from the large scale into the viscous/micro scale where the excess angular momentum is released as heat. In the most widely used approach, the so-called α-disk, the disk is treated as a set of nested rings with matter orbiting as per Kepler's laws. For the matter from the outer disk to reach the stellar surface, viscosity and shear between adjacent rings need to be invoked. The excess energy is transported to (and dissipated at) the eddy scale (i.e., the smallest scales of the turbulence), but angular momentum is transported to large radius, producing the subsequent heating of the disk. The spectral energy distribution predicted by this simple model fits well with the integrated radiative output from PPDs. in spite of the uncertainties about its physical foundations and operation. Since 1991, Magneto Rotational Instability (MRI) has been thought to play a key role in enabling the radial transport of mass. However, to operate, MRI requires disk ionisation fractions that are not met by the gas in some areas of PPDs; for solar like stars, this MRI dead zone is located in the planet formation zone (at radii from 1 to 10 AU). Many ideas have been advanced to solve this problem in the context of PPD evolution. To mention but a few: the onset of turbulent mixing in the PPDs dead zone, the action of gravitational instabilities and density waves as additional mechanisms for angular transport, or a more realistic evaluation of the energy budget irradiating the disk, including the outflow contribution. However, only direct imaging of the inner 20 AU of PPDs can provide the data to determine the sources of radiation accurately, to measure the radiation field reaching the surface of PPDs, map the gravitational/density waves acting on the disk and study the variation of the disk temperature from the mid plane (at IR wavelengths) to the atmosphere (at UV wavelengths) for radii in the 1-100 AU range. Milliarcsecond spatial resolution (MASI) is required to rigorously address the physics of PPDs since all spectroscopic means rely on the Keplerian rotation of the disk and derive the disk structure based on the Doppler shifting/broadening of the relevant spectral tracers.

## 4.2 The connection between PPDs and jets

Though bipolar outflows are the first signature of star formation, no theory anticipated them prior to the discovery of the outflow from L1551. The collimation and mechanical power of outflows decreases as stars approach the main sequence and protoplanetary disks (PPDs) evolve into debris disks. The tracers of the outflows depend on the degree of evolution of the parent source. Flows from very young sources ($10^4$-$10^5$ yrs old) often contain molecules such as $H_2$, CO, and SiO. Later on, they tend to be dominated by HI and low-ionization metals that are observed as prominent large-scale jets detected at optical wavelengths ([S II], [N II], or Hα lines). Mass-loss rates decline from ~$10^{-6}$ $M_\odot$ yr$^{-1}$ in the youngest sources to ~$10^{-10}$ $M_\odot$ yr$^{-1}$ before being switched off. Typically, the mass-loss rate is ~10% of the PPD accretion rate and the velocity about few hundred km s$^{-1}$.

Outflows are thought to be composed of:
- An episodic component ejected from the boundary layer between the star and the disk that carries on the excess magnetic energy generated by the magnetic coupling between the star and the inner border of the disk.
- A stable component or disk wind that carries on the toroidal magnetic field and keeps the wind collimated.

Required diagnostics that cannot be observed as today are:
- Rotation of the jet (if disk wind is magnetised and carries on a toroidal field it should be rotating)
- Connection between magnetized disk winds and photoevaporative flows
- Source of the episodic emission and connection with accretion
- Irradiation of the disk by the outflow.

Many uncertainties remain due to the lack of observations to constrain the modelling. Very important questions still open are: What role do disk instabilities play in the whole accretion/outflow process? What are the dominant processes involved in wind acceleration? How does this mechanism work, if it does, when radiation pressure becomes significant as for Herbig Ae/Be stars? What are the relevant time-scales for mass ejection? How does the high-energy environment affect the chemical properties of the disk and planetary assembly? How do the stellar magnetospheres evolve? What is the role of binarity and planet formation in disk evolution? MASI at UVOIR wavelengths is needed to address them.

## 4.3 Globular clusters and nearby stellar populations

The nearby Universe offers us a unique window to the processes of star formation and galaxy assembly. The depth and level of detail of nearby studies provide us with a reference to calibrate more distant systems, and represent the only opportunity to constrain and understand the underlying physics at the (sub)parsec level.

Stellar clusters were traditionally thought of as excellent examples of "simple stellar populations" - consisting of stars with a single age and a single chemical composition. While this still holds true for low-mass *open* clusters in the Solar neighbourhood, it is clear that massive *globular* clusters (GCs) that inhabit the Galactic halo are far more complex systems. Colour-magnitude diagrams from HST show *multiple populations* that reveal themselves through parallel main sequences, split red giant branches, and other features (Milone et al., 2017). In massive GCs in the Milky Way, multiple populations are associated with enhancements in light elements and helium content (Gratton et al. 2012, Bastian & Lardo 2018).

Two key capabilities of HST have been crucial in uncovering the variety of this phenomenon in old GCs: 1) from space, it is possible to achieve exquisite photometric accuracy that is very hard to reach from the ground, especially in crowded environments, 2) space-based observations provide access to important spectral features in the UV (CH, CN, and NH molecular bands) that are sensitive to the light-element abundance variations that trace the multiple populations (see Fig. 10). It is important to note that none of the currently planned future facilities (JWST, Euclid, E-ELT…) will deliver improvements in these two critical aspects simultaneously.

*Young* star clusters with globular cluster-like masses ($>10^5$ $M_\odot$) are rare and tend to be located beyond the Local Group. Even with HST, it is extremely challenging to obtain photometry of individual stars in such clusters, and only crude information exists about their stellar contents (Larsen et al. 2011). Imaging with a 10 m space telescope will, however, make it possible to obtain exquisite photometry for hundreds of post-main sequence stars per cluster, as well as thousands of fainter stars. In addition to providing unprecedented constraints on stellar evolution (e.g. the role of binary/multiple stars and stellar rotation), such observations would also provide a key constraint on formation scenarios for multiple populations by establishing (or ruling out) their existence in young clusters (and other environments, such as galaxy nuclei), establishing whether age spreads are present, and determining the fractions of stars belonging to the different populations.

The GC systems of galaxies beyond the LG offer unique complementary diagnostic power to the problem of multiple populations (Bastian & Lardo 2018). A 10m-class telescope in space will provide ~1pc resolution and unprecedented sensitivity (AB~31mag) out to a distance of ~20Mpc. This volume includes the two closest galaxy clusters, Virgo and Fornax. These environments represent the natural habitats of compact stellar systems, harbouring *tens of thousands* of old GCs and ultra-compact dwarfs (Durrell et al. 2013). A 10m-class telescope in space will be able to infer the existence of such He-enriched subpopulations through accurate UV-to-optical colours, because a UV excess is characteristic of HBs that are well populated at high effective temperatures. For the first time, it will be possible to fully resolve these compact stellar systems radially, further constraining the relative radial contribution of the subpopulations. These studies will finally assess whether the presence of helium-enriched subpopulations is a widespread phenomenon of all GCs, or if it depends on the nature (stellar mass, morphological type) of the host galaxy.

Unfortunately, detailed studies of field stars and stellar clusters are mostly confined to the Local Group (LG) because of sensitivity, resolution and crowding effects. While much has been learned of the ages and elemental abundances of stars in LG galaxies, cosmic variance still constitutes a fundamental limitation. A 10m-class observatory with UV-to-NIR imaging capabilities in space will allow the study of the star formation and chemical enrichment histories, the dust content, the massive end of the stellar mass function, and the populations of star clusters of all galaxies within the Local Volume (~10 Mpc). This will represent an order of magnitude increase in sample size, finally including galaxies of all luminosities, morphological types, and inhabiting a variety of environments. Access to UV imaging at high spatial resolution is of particular relevance to address questions related to star clusters and hot stars, and when combined with optical and NIR photometry allows for simultaneous determination of effective temperatures and extinction.

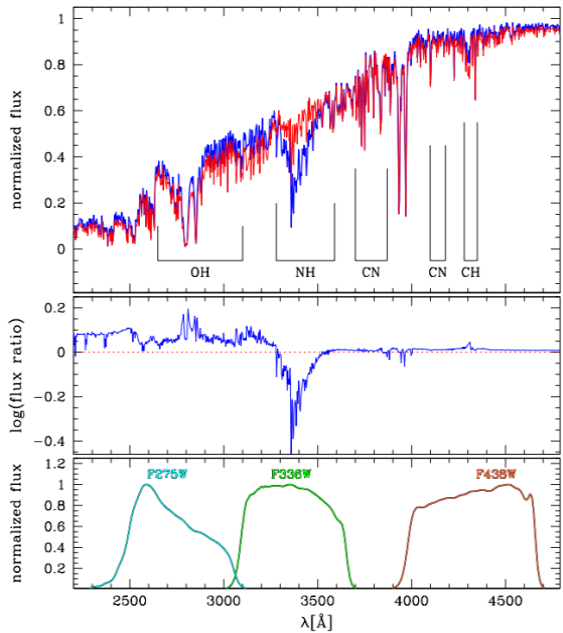

*Figure 10. The top panel shows a comparison between the simulated spectrum of a first-generation, N-poor RGB star (red) and a third generation, N-enriched one (blue). The middle panel shows the flux ratio of the two spectra, where it is clear that the signature of multiple populations is confined to wavelengths λ < 450nm. This is why the UV capabilities of the WFC3/UVIS camera onboard HST (bottom panel) have been instrumental in the discovery and characterization of this phenomenom. From Piotto et al. (2015).*

## 4.5 Massive stars

Massive stars are cosmic engines and probes of the Universe. In life they are mighty sources of ionising radiation and mechanical energy, leaving their striking imprint as bubbles and giant HII regions or as galactic-scale outflows. In death they are the progenitors of supernovae and long γ-ray bursts, events so bright they can be detected in distant galaxies, offering promising potential probes of star formation up to redshift z=9 (Robertson & Ellis, 2012). During the explosion a number of chemical elements essential to life as we know it are released, the most important of which is Oxygen. Their remnants are neutron stars, pulsars and black holes, sites of extreme physics that will guide us through the Universe in the new domain of gravitational waves opened by LIGO and Virgo (e.g. Abbott et al., 2016).

Insights into these topics require an excellent understanding of the lifecycle of massive stars, including their physical properties and evolutionary sequence, both of which are strongly affected by their chemical compositions. Massive stars in the low-metallicity regime are key to interpreting the Universe as we look toward the earliest times, particularly toward the near metal-free 'first stars' (Population III). However, current facilities put severe limits on our ability to obtain robust empirical constraints at the lowest metallicities. The SMC (at 60 kpc) is the most metal-poor environment where a broad sample of massive stars can be resolved and analyzed in detail today, forming the current standard for the metal-poor regime, e.g. spectral libraries of its massive stars feed population-synthesis codes that model the far Universe such as Starburst99 (Leitherer et al. 2014). However, the SMC's $1/5\ Z_\odot$ metallicity is far from the $1/300\ Z_\odot$ detected in quasars at z=5 (Prochaska et al., 2003).

A vital aspect to our understanding of the evolution of metal-poor massive stars is their stellar winds. During most of their lives, massive stars shed mass by the action of their own radiation field at a rate that decreases with decreasing luminosity and metallicity (Kudritzki, 2002). This prediction has been confirmed observationally down to the metallicity of the SMC (Mokiem et al., 2007) and, indeed, radiation-driven mass loss is considered negligible in models of Pop III stars (Marigo et al., 2003; Ekstrom et al., 2008). This point is relevant as the mass-loss prescription in the models of stellar evolution can significantly change the evolutionary sequence, stellar feedback, and the final pre-supernova mass.

Ground-based 8-10m telescopes enabled first exploration of galaxies beyond the SMC with lower oxygen abundances: IC1613, WLM and NGC3109, finding unexpected results such as an LBV with strong P Cygni profiles (Herrero et al. 2010), an extreme oxygen-rich W-R star (Tramper et al. 2013), and indications of stronger-than-expected stellar winds from analysis of visible spectroscopy (Herrero et al. 2012, Tramper et al. 2014). The essential need for UV observations was demonstrated by *HST*-COS observations of OB stars in IC1613 which showed that their wind strength had been overestimated (Bouret et al., 2015; Garcia et al., 2014). Only with the UV data were we able to glean that the sample stars are oxygen-poor, but their iron content is roughly comparable to that of the SMC. In other words, the truly metal-poor regime remains unexplored.

The evolutionary sequence of metal-poor massive stars is also yet to be defined. Very fast rotating stars could undergo a special channel, called chemically homogeneous evolution (CHE) where they stay compact with high effective temperatures (Szecsi et al., 2015), leading to increased ionizing fluxes by factors of two to four. This is one of the channels invoked to explain the binary systems leading to the gravitational wave signals detected by LIGO (de Mink & Mandel, 2016). However, we have been unable to test the predictions of CHE due to the paucity of observations in sub-SMC metallicity environments.

*HST has produced outstanding contributions to the advancement of massive stars, but it is time for a sizeable step forward.* IC1613, WLM, NGC3109 and the promising $1/10Z_\odot$ galaxy Sextans A (Camacho et al., 2016) are located at 0.7-1.2 Mpc, i.e., a factor 10 to 20 further than the SMC and at the limit of current facilities. Two other extremely metal-poor star-forming galaxies have been recently surveyed for O-type stars, SagDIG ($1/20Z_\odot$; Garcia 2018) and Leo P ($1/30Z_\odot$; Evans et al. 2019), but they have relatively sparse populations as well as the fact that their distance and foreground extinctions limit optical (let alone UV) follow-up. In short, greater sensitivity combined with excellent spatial resolution (at visible and UV wavelengths) is needed to enable the required quantitative studies in the metal-poor regime. We also note that only UV spectroscopy can provide insights on the stratification of stellar winds and their inhomogeneities (Sundqvist et al., 2014; Fullerton et al., 2006). UV access, together with ALMA and/or ATHENA is essential to characterise radiation-driven winds in extragalactic massive stars.

A formidable facility such as LUVOIR will allow us to study individual stars in DDO68 and IZw18, an important landmark in the road to the low-metallicity Universe with $1/32Z_\odot$ (Vilchez & Iglesias-Paramo 1998; Annibali et al. 2019). Located at distances of 13 and 18 Mpc, respectively, their populations cannot be resolved by current instrumentation, but would be by LUVOIR working at the diffraction limit in the UV. LUVOIR would also open-up observations of resolved massive stars in a diverse range of galaxies out to several Mpc, including galaxies in the Sculptor, Centaurus and M81 groups (see e.g. KINGFISH: Kennicutt et al., 2011).

Reconstructing the lifecycle of metal-poor massive stars requires observations of quasi-coeval populations and determination of physical properties of large samples. This is typically performed in three stages:

- *Photometric census:* High quality optical/near-IR photometry is crucial to study the star-formation history of external galaxies and to identify candidate massive stars. *HST* has proved superior to larger ground-based telescopes in this respect, with minimal sky background levels and a spatial resolution unmatched by adaptive optics at the shorter wavelengths, providing outstanding photometric catalogs of resolved populations in Local Group and nearby galaxies (e.g. Tolstoy et al. 2009; ANGST: Dalcanton et al. 2009, HTTP: Sabbi et al. 2013). The enhanced collecting power of LUVOIR will comfortably reach IZw18, and continue the search of very massive stars in gas- and dust-enshrouded nurseries of closer galaxies. Its spatial resolution will be key to push such studies of resolved populations beyond the Local Group.

- *Multi-object spectroscopy:* Visible/near-IR spectroscopy is then needed to confirm the candidate stars and determine their physical properties. Because of the large integration times involved, this is typically performed with multi-object spectrographs from the ground. The ELT will be able to reach the brightest OB stars out to ~4 or 5 Mpc. However, panchromatic UV to near-IR spectra with stable flux calibration from space are also needed, as only SED fitting can yield extinction towards the star, hence accurate luminosities and stellar masses. Moreover, at visible wavelengths the ELT will significantly limited by the effects of Earth's atmosphere – high-performance adaptive optics at blue wavelengths is still a long way in the future, space-borne spectroscopy with a LUVOIR-like platform will substantially out-perform ELT spatially, and be competitive in terms of overall sensitivity.

- *UV spectroscopy:* UV observations will enable characterisation of the winds of metal-poor massive stars. The 1200-1800 Å range, accessible only from space, holds exclusive information on the wind velocity and is the spectral range were hot massive stars emit the most. Sparing regions of severe extinction, both LUVOIR concepts will comfortably reach the faintest O-stars out to distances of several Mpc (see Figure 11). Pushing the facility to its limits, and thanks to the spatial resolution improvement, it will also reach the brightest OB stars in galaxies such as DDO68 and IZw18, providing a unique chance to investigate the properties of massive-stars in near primordial conditions.

A more expansive version of this case is given in the Voyage2050 White Paper by Garcia et al.

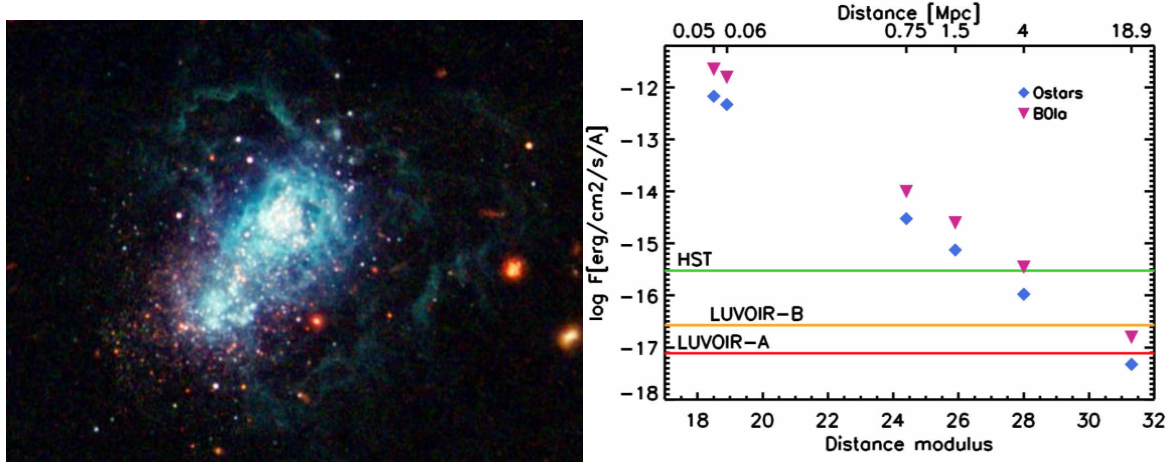

*Figure 11. Left: With a metal content of 1/32$Z_\odot$, IZw18 represents the next step down in the metallicity ladder towards the metal-free early Universe. IZw18 has a demonstrated content of massive stars, but being ~18Mpc away current facilities cannot resolve its stellar population. Right: Expected UV fluxes at 1500Å for OB stars located at successively farther galaxies, compared to the performance of HST and the two LUVOIR architectures. The stellar UV fluxes are typical values found in IC1613 scaled by distance. The plotted sensitivity of HST represents the faintest flux for which 5 orbits grant a spectrum with sufficient quality for quantitative analysis; this flux was scaled by mirror surface to estimate how deep LUVOIR would reach. LUVOIR-A will enable breakthrough spectroscopy of individual bright OB stars in IZw18.*

## 5 Galaxy evolution

A large UVOIR telescope affords an opportunity for remarkable progress in the study of galaxy evolution, both through study of distant systems and via 'galactic archeology' of more nearby systems. A programme which takes advantage of the likely capabilities of such a mission, including high-resolution imaging, especially in the UV, much higher contrast and lower background levels, and a stable PSF over several arcminutes, can carry out studies which will not be possible either with *JWST* or with the next generation of large ground-based telescopes (e.g. E-ELT). While much progress has been made in understanding galaxy formation, many important questions remain unanswered, and in need of progress in instrumentation. Studies carried out with HST have shown that galaxies observed during the epochs when star formation was most active and when galaxies were still assembling look different to today, but lack the resolution to study the processes which shape this morphological difference. Small and dense systems somehow form the galaxies we see today; the processes responsible for disk growth in particular remain obscure, and the effect of the many different feedback mechanisms on the galaxy population remains frustratingly difficult to pin down.

The sensitivity, spectroscopic resolution and image quality of LUVOIR at UV and optical wavelengths will open a new observational window on the assembling galaxy population observed at the cosmic times corresponding to the peak of star formation activity in the Universe, i.e. around redshifts 1<z<6. It will complement and extend the results expected in the coming years by JWST, which will enable the discovery of the redshifted UV and optical emission of the first sources that illuminated and re-ionized the Universe at redshift z>6, by observing at near-infrared wavelengths (0.6–28µm) with superb angular resolution (60 mas at 2µm) and medium-resolution spectroscopic capabilities (R~2700). The superior sensitivity and image quality of LUVOIR will dramatically improve the exploration of this population at observed wavelengths below ~1–2µm, where the deepest observing campaigns with the much less sensitive HST have led to the photometric detection of only modest numbers of galaxies.

The outstanding sensitivity foreseen for LUVOIR will be particularly crucial to address a number of fundamental scientific questions that even JWST will not be able to solve. For instance, its unprecedented UV-optical sensitivity will allow us to establish which sources were responsible for the re-ionisation of the neutral intergalactic medium, by directly measuring the leakage of ionising photons produced by massive stars and accreting black holes from the first galaxy sub-units. Similarly, high-spectral resolution capabilities in the UV will enable for the first time spatially resolved spectral analyses of even faint galaxies in the first epoch of assembly, by measuring the metallicity gradients, the kinematic structure and the ionization structure with unprecedented sub-kpc resolution. This will exploit the unique capability of LUVOIR of measuring the emission-line spectrum at rest-UV wavelengths, which are far more informative than rest-optical lines to constrain the early chemical enrichment of galaxies by different types of supernovae and the mapping of galactic outflows (and the enrichment of the intergalactic medium).

To achieve these ambitious scientific goals, it is important to equip LUVOIR with a suitable range of instruments capable of high resolution, spatially resolved spectroscopy - either large field IFUs or multi-IFUs. These UV-sensitive instruments will allow one to resolve individual star-forming regions in young galaxies and directly measure the relative importance of the various processes that are expected to shape galaxy formation: star formation, nuclear activity by an accreting black hole, and interactions/mergers with satellites. Wide-field imaging capabilities and spectroscopic multiplexing will allow such studies in cosmological volumes. The combination of a large UVOIR telescope (delivering spatially resolved UV-vis data) and the next generation of ground-based ELTs (securing the required long-wavelength information after the end of the JWST mission) will yield a unique view of the physics of high redshift galaxies.

## 6 Scientific Requirements - Exoplanets

Each science topic has a range of science requirements, direct imaging and spectroscopy of exoEarths being the greatest challenge in terms of signal-to-noise, spatial resolution and imaging contrast. Figure 12 shows estimates of the number of Earth-size planet detections in stellar habitable zones as a function of telescope aperture. A mirror diameter in excess of 8m is required to begin to supply statistically interesting numbers of potential targets. The requirements of a range scientific topics are summarized below.

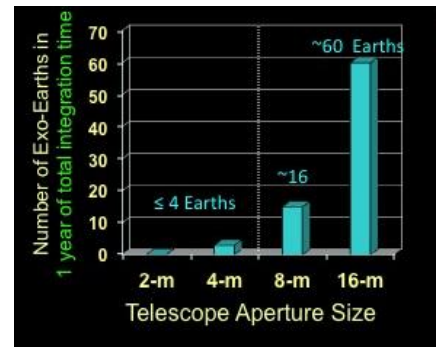

### 6.1 Direct imaging

To image an Earth twin (1 AU from its star) at a distance of 20 pc requires achieving a star-planet contrast of $10^{10}$ at 50 mas from the star. All three configurations considered in the table achieve this milestone, if just barely. The 12 m and 16 m variants would allow imaging of an Earth twin with an internal coronagraph out to 30 pc and 40 pc respectively.

*Figure 12. Estimated number of Earth-like planets around long-lived stars as a function of telescope aperture. The calculation assumes 10% of Solar type stars have Earth-like planets in their habitable zones.*

The Search for Living Worlds – ESA Voyage 2050 White Paper

| Observation Requirement | Major Progress | Substantial Progress | Incremental Progress |
|---|---|---|---|
| Wavelengths | 0.4 – 2 µm with internal coronagraph, 2.5 – 5 µm with next-gen Mid-IR imager | 0.4 – 2 µm | 0.4 – 2 µm |
| Spectral Resolution | R~100 | R~40 | R~40 |
| Field-of-View | 10" | 10" | 10" |
| Inner Working Angle | 25 mas (3 λ / D @ 650 nm with internal coronagraph), 8 mas (1 λ / D @ 650 nm with starshade) | 35 mas (3 λ / D @ 650 nm) | 50 mas (3 λ / D @ 650 nm) |
| Contrast | $10^{10}$ at IWA | $10^{10}$ at IWA | $10^{10}$ at IWA |
| Telescope Aperture | 16 m | 12 m | 8 m |
| Coronagraph Type(s) | Internal coronagraph and starshade | Internal coronagraph | Internal coronagraph |

## 6.2 Atmospheric Escape

| Observation Requirement | Major Progress | Substantial Progress | Incremental Progress |
|---|---|---|---|
| Wavelengths | 100 – 1000 nm | 100 – 1000 nm | 100 – 1000 nm |
| Spatial resolution | 0.1" | 0.1" | 0.1" |
| Spectral resolution | UV: 60,000 Optical: 100,000 | UV: 30,000 Optical: 70,000 | UV: 20,000 Optical: 50,000 |
| Field-of-view | N/A | N/A | N/A |
| Contrast | N/A | N/A | N/A |
| Telescope aperture | 16m | 8m | 6m |
| Exposure time | 1 – 30 h / target | 1 – 15 h / target | 1 – 8 h /target |
| Time resolution | 1s | 10s | 20s |

## 6.3 The bulk composition of rocky planets

| Observation Requirement | Major Progress | Substantial Progress | Incremental Progress |
|---|---|---|---|
| Wavelengths | 91 – 320 nm | 91 – 320 nm | 91 – 320 nm |
| Spectral resolution | 50,000 | 35,000 | 20,000 |
| Telescope aperture | 16m | 8m | 6m |
| Exposure time | 0.5h / target | 2h / target | 4 h /target |

## 6.4 Transiting exoplanets

| Observation Requirement | Major Progress |
|---|---|
| Wavelengths | 0.1-2 microns (up to 5 microns ideal) |
| Spectral resolution | >100 (1000 ideal) |
| Coronograph | 0.4-2 microns (up to 5 better) R~100 or better |

The Search for Living Worlds – ESA Voyage 2050 White Paper

Extension of the wavelength coverage to 5 μm would be particularly useful for exoplanet science (Werner et al. 2016), as it would increase the opportunity to detect thermal signatures from hot exoplanets. Transit spectroscopy signatures are very small, of order 1:10,000 to 1:1000,000, so maximising the number of photons collected is necessary. Provided systematic noise remains constant, larger mirror sizes are most advantageous for this kind of observation.

## 7. Possible telescope solutions

The fundamental difficulty of analysing distant worlds is their intrinsic faintness. The broadband reflected light from an exoEarth is not expected to be much more than 10 nanojansky. This is further complicated by their proximity to their host stars, which are typically tens of millions to tens of billions times brighter. These challenges lead to key requirements for making such direct measurements of the composition of the atmospheres of Earth-sized exoplanets. We need to be able to collect enough photons from the planet so that we can obtain a low-resolution (R ~ 70–100) spectrum with a good signal-to-noise ratio (SNR ~ 10) over wavelength ranges that include key habitability and bio-signature gas spectral features, many of which occur in the UV and visible ranges. Other key drivers for a telescope to address this science goal are to be able to resolve an object within the habitable zones of stars out to ~20pc and suppress the overwhelming scattered light from the host star. Observing in the visible and UV confers a clear benefit for the diffraction limit. Characterising dozens of exoEarth candidates is a minimum scientific goal.

This requires:
- Large aperture telescope of at least 10-12m in diameter
- Angular resolution of 10 mas
- Starlight suppression from a coronagraph with contrast ~$10^{-10}$
- Ultra-low sky background
- Access to the UV

The combination of these stringent requirements can only be achieved using a telescope flown in space, in a deep L2 (or similar) orbit. The likely configuration will build on the approach developed by JWST, as illustrated in figure 13. While the requirements on the telescope size and location are set by the planetary science, we envisage a broad range of instruments to support a broad science programme, mirroring the approach of HST and JWST. A project of this magnitude can and should serve the whole astronomical community by providing observatory type facilities, with a suite of 3-4 instruments optimised to give the maximum scientific output. More detailed requirements on wavelength coverage and spectral resolution are summarized in the tables in section 6.
- Coronographic imager and low-resolution spectrograph
- High resolution UV spectrograph with coronographic capability
- Wide field imager
- Polarisation analyser (to be included in spectrographs)
- Multi-object spectrograph

Trade-offs between the various instrument requirements will need to be carried out as part of detailed studies to consider technology needs and optimise instrument designs.

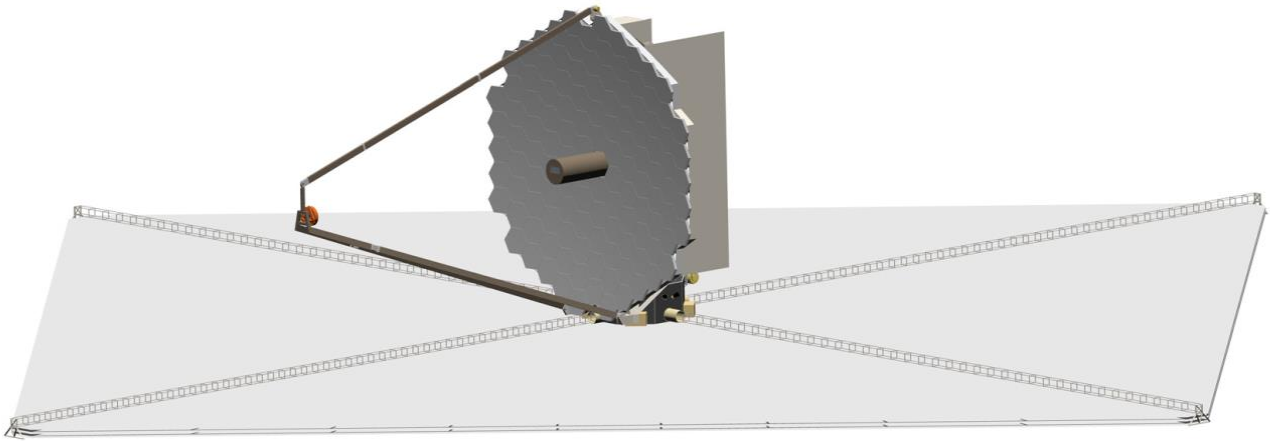

*Figure 13. Preliminary rendering of the LUVOIR Architecture A observatory, which has a 15-meter diameter on-axis primary mirror and four instrument bays. An animation of the observatory deployment and pointing may be viewed at* https://asd.gsfc.nasa.gov/luvoir/design/. *Credit: A. Jones (NASA GSFC)*

Due to the likely cost, a large UVOIR telescope will need to be a multi-agency mission combining at least two partners. While it will build on significant heritage from HST and JWST, there will be significant development programmes required to reach minimum TRL levels for the subsystems by the mid-2020s, to fly a mission in the mid-2030s.

For some of the key technologies required for the future large UVOIR space observatory, as Silicon Carbide for structure and lightweight mirrors (ADS/Boostec), CCD detectors systems (e2V) or MCP detectors (Photonis), the principle suppliers are European companies. Beside this industrial leadership, significant European expertise in several subsystems and key technologies, based on developments pursued by agencies, laboratories or industry, also provides an important opportunity for ESA, including:

- Space active optics (Deformable mirrors and wavefront sensing
- Freeform optics manufacturing and innovative optical design (compactness, very wide-field, ..)
- Coronographic wavefront sensors (both in pupill and focal plane)
- Cophasing wavefront sensors,
- Curved detectors and innovative focal plane arrays
- Programmable slits using µmirror array (MOS)
- Programmable spectrograph (in wavelength) using µmirrors

## 8. Relevance of LUVOIR to ESA

There is significant interest in the development of a large UV-optical-IR telescope in the USA. During the past few years, studies have taken place around the concept of an Advanced Technology Large Aperture Space Telescope (ATLAST) and an AURA commissioned a detailed report entitled "From Cosmic Birth to Living Earths" focused on the definition of a High Definition Space Telescope (HDST). Currently, NASA is carrying out four studies of flagship large missions in preparation for the next Decadal Survey of astronomy. One of these is a Large UV-Optical-IR space telescope (LUVOIR). Several European scientists (e.g. M. Barstow, Ferrari, Gomez de Castro, Henning, Buchhave) act as observers on the LUVOIR Science and Technology Definition Team, nominated by their national space agencies.

While the LUVOIR study is very much a US-focused look at the next potential NASA Flagship astronomy mission, there are clear interests in Europe. A 34-strong team of scientists (core team listed below) has prepared this white paper with the support of a much wider community. Furthermore, the costs of such a mission will very likely require NASA to seek partners that can make a substantial contribution to the project. We propose that, if it selected, **ESA should join the LUVOIR project with and M or L-class contribution**, to the benefit the European exoplanet research and the astronomical community at-large.

The Search for Living Worlds – ESA Voyage 2050 White Paper

Lead Proposer - Martin Barstow, Department of Physics & Astronomy, University of Leicester, University Road, Leicester, LE1 7RH, UK. email: mab@le.ac.uk, tel: +44 116 252 3492

Core Proposing Team – S. Aigrain (U. Oxford, UK), J. Barstow (UCL, UK), M. Barthelemy (U. Grenoble, France), B. Biller (ROE, UK), A. Bonanos (National Observatory Athens, Greece), L. Buchhave (U. Copenhagen, Denmark), S. Casewell (U. Leicester, UK), C. Charbonnel (U. Geneva, Switzerland), S. Charlot (IAP, France), R. Davies (U. Oxford, UK), N. Devaney (U. Galway, Ireland), C. Evans (ATC, Edinburgh, UK), M. Ferrari (Laboratoire d'Astrophysique, Marseille, France), L. Fossatti (IWF, Austria), B. Gänsicke (U. Warwick, UK), M. Garcia (CAB-CSIC, Spain), A. Gomez de Castro (U. Complutense, Madrid, Spain), T. Henning, (MPIA, Germany), C. Lintott (U. Oxford, UK), C. Knigge (U. Southampton, UK), C. Neiner (LESIA, OP, Paris, France), L. Rossi (U. Delft, The Netherlands), C. Snodgrass (U. Edinburgh, UK), D. Stam (U. Delft, The Netherlands), E. Tolstoy (U. Groningen, The Netherlands), M. Tosi (INAF, Bologna, Italy).

# Bibliography

Abbott, B.P., et al. 2016, Phys.Rev. Lett., 116, 061102
Agócs et al, 2014, SPIE
Anglada-Escudé et al. 2016, Nature, 536, 381.
Annibali et al., 2019, MNRAS, 482, 3892
Barstow, J. K. et al., 2014, ApJ, 786, 154
Bastian, N., & Lardo, C., 2019, ARAA, 56, 83
Ben-Jaffel & Ballester 2013, A&A, 553, A52
Bétrémieux, Y. & Kaltenegger, L., 2014 ApJ, 791, 7.
Bieler, A., et al., 2015, Nature, 526, 678
Bisikalo et al. 2013, ApJ, 764, 19
Bockelée-Morvan, D. et al. 2015, Space Science Reviews, 197, 47-83
Bonnefoy, M. et al., 2016, Astronomy & Astrophysics, 587, 58.
Bouret, J.-C., et al. 2015, MNRAS, 449, 1545
Bourrier et al. 2016, A&A, 591, A121
Bourrier et al. 2015, A&A, 573, A11
Camacho, I., et al. 2016, A&A, 585, A82
Canonica, et al., 2013, *J. Micromech. Microeng.* 23, 055009
Carter-Bond, J. C., et al., 2012, ApJ, 747, L2
Chadney et al. 2015, Icarus, 250, 357
Cockell et al. 2016, Astrobiology, 16, 89
Crossfield, I.J.M, 2013 Astronomy & Astrophysics, 551, 99.
Crossfield, I.J.M. et al., 2014, Nature, 505, 654.
Crossfield, I.J.M., 2014, Astronomy & Astrophysics, 566, 130.
Dalcanton, J. J., et al. 2009, ApJS, 183, 67
Davidsson, B. et al. 2016, A&A, 592, A63
Debes, J. H. & Sigurdsson, S. 2002, ApJ, 572, 556
Demory et al. 2016, Nature, 532, 207
Domagal -Goldman, S. D. et al., 2014, ApJ, 792, 90.
Denolle et al, 2013, SPIE, 8713
Dinyari et al, 2008, Appl. Phys. Lett. 92, 091114
Dorn, C., et al. 2015, A&A 577, A83
Dumas et al., 2010, SPIE. 7742
Ehrenreich et al. 2015, Nature, 522, 459
Ekström, S., et al. 2008, A&A, 489, 685
Evans et al. 2019, A&A, 622, A129
Evans, T. M. et al., 2013, ApJL, 772, L16
Feldman, P. et al. 2015, A&A, 583, A8
Fendler et al., 2012, SPIE, 8453
Fletcher, L. N. et al., 2011, Icarus, 214, 510
Fossati, L. et al. 2010a, ApJL, 714, L222
France, K. et al. 2018, ApJS, 239, 16
Fullerton, A. W., et al. 2006, ApJ, 637, 1025
Gänsicke, B. T., et al. 2012, MNRAS, 424, 333
Garcia et al. 2014, ApJ, 788, 64
Garcia, 2018 MNRAS, 474, L66
Giles, R. S., Fletcher, L. N. & Irwin, P. G. J., 2015, Icarus, 257, 457
Gillon, M. et al., 2016, Nature, 533, 221
Gillon, M. et al., 2017, Nature, 542, 456
Gladstone, G. R. & Yung, Y. L., 1983, ApJ, 266, 415
Gratton, R. G., et al., 2012, A&A Rev., 20, 50
Hansen & Houvenier, 1974, JAtS, 31, 1137

The Search for Living Worlds – ESA Voyage 2050 White Paper

Harman, C. E. et al., 2015, ApJ, 812, 137.
Heap, S. R., et al. 2011, BSRSL, 80, 149
Herrero et al. 2010, A&A, 513, A70
Herrero et al. 2012, A&A, 543, A85
Hsieh, H., & Jewitt, D. 2006, Science, 312, 561
Ito et al. 2015, ApJ, 801, 144
Itonaga et al, 2014, Sony R&D Platform, Atsugi, Japan. Symposium on VLSI Technology,
Iwert O. and Delabre B., 2010, SPIE, 7742
Janin-Potiron et al, A&A 2016
Jenkins, J. et al., 2015, AJ, 150, 56.
Jewitt, D., et al., 2015, p.221-241, in "Asteroids IV", Michel, P., DeMeo, F., & Bottke, W. (eds.), University of Arizona Press
Jin et al. 2014, ApJ, 795, 65
Jura, M. 2003, ApJ 584, L91
Karalidi et al., 2012, A&A, 546, 56
Karkoschka, E. & Tomasko, M., 2005, Icarus, 179, 195
Kataria, T. et al., 2015, ApJ, 801, 86
Kennicutt, R. C., et al. 2011, PASP, 123, 1347
Kehrig, C., et al. 2015, ApJL, 801, L28
Keller et al., 2019, SPIE, 7735, 6
Kemp et al., 1987, Nature, 326, 270
Khodachenko et al. 2015, ApJ, 813, 50
Kislyakova et al. 2014, Science, 346, 981
Kislyakova et al. 2016, MNRAS, 461, 988
Knutson, H. A. et al., 2012, ApJ, 754, 22
Ko, et al., 2008, Nature, 454, 748
de Kok et al., 2011, ApJ, 741, 59
Koskinen et al. 2010, ApJ, 723, 116
Kreidberg, L. et al., 2014, Nature, 505, 69.
Kubyshkina, D., et al. 2018, A&A, 619, A151
Kubyshkina, D., et al. 2019, ApJ, 879, 26
Kudritzki, R. P., 2002, ApJ, 577, 389
Kulow et al. 2014, ApJ, 786, 132
Kurokawa & Kaltenegger 2013, MNRAS, 433, 3239
Lammer et al. 2009, A&ARv, 17, 181
Larsen, S. S., et al. 2011, A&A, 532, A147
Lawson, P. et al., 2012, Proc. SPIE, 8447, 22
Lecavelier des Etangs et al. 2004, A&A, 418, L1
Lecavelier des Etangs 2007, A&A, 461, 1185
Lee, J.-M. et al., 2014, ApJ, 789, 14
Leitherer et al. 2014, ApJS, 212, 14
Leitzinger et al. 2011, A&A, 536, A62
Line, M. R. et al, 2014, ApJ, 783, 13
Line, M. R. & Parmentier, V., 2016, ApJ, 820, 78
Linsky et al. 2014, ApJ, 780, 61
Lopez & Fortney 2013, ApJ, 776, 2
Luger, R. and Barnes, R., 2015, Astrobiology, 15,2, 119-143.
Macintosh, B. et al., 2015, Science, 350, 64.
Marigo, P., et al. 2003, A&A, 399, 617
Marley, M. et al., 2014, arXiv1412.8440
Mawet, D. et al., 2012, Proc. SPIE, 8442, 4
Miguel et al. 2011, ApJL, 742, L19
Milone, A., et al. 2017, MNRAS, 464, 3636
de Mink, S. E., & Mandel, I. 2016, MNRAS, 460, 3545
Mokiem, M. R., et al. 2007, A&A, 473, 603

Morley, C. et al., 2015, ApJ, 815, 110
Moriarty, J., et al., 2014, ApJ, 787, 81
Mura et al. 2011, Icarus, 211, 1
N'Diaye et al, A&A 2016
Nikzad, S. et al, 2006, JPL, SPIE,
Owen & Jackson 2012, MNRAS, 425, 2931
Paul et al, A&A 2014
Pont, F. et al., 2013, MNRAS, 432, 2917
Pope et al, A&A 2016
Prochaska, J. X., et al. 2003, ApJL, 595, L9
Rappaport et al. 2012, ApJ, 752, 1
Robertson, B. E., & Ellis, R. S. 2012, ApJ, 744, 95
Rogers, L. A. & Seager, S. 2010, ApJ, 712, 974
Rossi, L. & Stam, D., 2018, A&A, 607, A57
Rubin, M., et al., 2015, Science, 348, 232
Rubio-Díez, M. M., et al. 2016, in prep
Rugheimer, S. et al., 2015, ApJ, 809, 57.
Sabbi, E., et al. 2013, AJ, 146, 53
Schwartz, J. C. & Cowan, N. B., 2015, MNRAS, 449, 4192
Seager et al., 2000, ApJ, 540, 504
Segura, A., et al., 2005, Astrobiology, 5,6, 706-725.
Sing, D. et al., 2016, Nature, 529,59.
Skemer, A. et al., 2012, Astrophysical Journal, 753,14.
Sing, D. K. et al., 2016, Nature, 529, 59
Skemer, A. et al., 2014, Astrophysical Journal, 792,17.
Stam et al., 2004, A&A, 428, 663
Stevenson, K. B. et al, 2014, Science, 346, 838
Sundqvist, J. O., et al. 2014, A&A, 568, A59
Swain, et al, 2004, SPIE 5301, 109
Szècsi, D., et al. 2015, A&A, 581, A15
Tekaya, K. 2014, Ph.D thesis, "Curved infrared focal plane array: hemispherical forming and induced optoelectronic properties".
Thompson & Rolland, 2012, Optics and Photonics news, 23, 6
Tolstoy, Hill & Tosi, 2009, ARAA, 47, 371
Tramper, F., et al. 2011, ApJL, 741, L8
Tramper et al. 2013, A&A, 559, A72
Tramper et al. 2014, A&A, 572, A36
Turbet, M. et al, 2016, A&A, 596, A112
Turner et al. 2016, MNRAS, 458, 3880
Vanderburg, A., et al. 2015, Nature, 526, 546
Vidal-Madjar et al. 2003, Nature, 422, 143
Vidal-Madjar et al. 2004, ApJ, 604, L69
Vidotto et al. 2010, ApJL, 722, L168
Vidotto et al. 2011, MNRAS, 414, 1573
Vidotto et al. 2013, EPJWC, 64, 04006
Viard et al, 2016, SPIE, 9906,
Veras, D. & Gänsicke, B. T. 2015, MNRAS, 447, 1049
Veras, D., et al., 2013, MNRAS, 431, 1686
Vílchez, J. M., & Iglesias-Paramo, J. 1998, ApJ, 508, 248
Wagner, K. et al., 2016, Science, 353,673.
Wihelm et al, 2008, Applied Optics, 47, 5473
Zamkotsian, et al., 2013, *SPIE*, 861618
Zuckerman, B., et al. 2007, ApJ, 671, 872